\documentclass[11pt,fleqn,leqno]{article}
\usepackage{graphicx}
\usepackage{amsfonts}
\usepackage{amsbsy}
\usepackage{latexsym}
\newcommand{\be}{\begin{equation}}
\newcommand{\ee}{\end{equation}}
\newcommand{\ba}{\begin{array}}
\newcommand{\ea}{\end{array}}

\newcommand{\Cp}{{\mathbb C\rm I\!P}}  
\newcommand{\Co}{{\mathbb C}}      
\newcommand{\s}{\sigma}
\renewcommand{\v}{{\mathsf{v}}}
\newcommand{\V}{{\mathsf{V}}}

\newcommand{\pr}{\prime}
\newcommand{\bt}{\beta}
\newcommand{\de}{\delta}

\newcommand{\bea}{\begin{eqnarray}}
\newcommand{\eea}{\end{eqnarray}}

\newcommand{\ga}{\gamma}

\newcommand{\al}{\alpha}

\newcommand{\vr}{\varrho}

\newcommand{\vp}{\varphi}
\begin{document}
\newtheorem{pro}[thm]{Proposition}
\newtheorem{lem}[thm]{Lemma}
\newtheorem{cor}[thm]{Corollary}
\newcommand{\C}{{\mathsf{C}}}
\newcommand{\U}{{\mathsf{U}}}
\newcommand{\A}{{\mathsf{A}}}
\newcommand{\G}{{\mathsf{\Gamma}}}
\renewcommand{\a}{{\mathsf{a}}}
\title{Eigenvalue correlations on Hyperelliptic
Riemann surfaces.}
\author{Yang Chen$^{\dag}$ and Tamara Grava$^{\dag\dag}$\\
Department of Mathematics\\
Imperial College\\
180 Queen's Gate, London, SW7 2BZ, UK}

\maketitle
\begin{abstract} 
\noindent
In this note we compute the functional derivative of the 
induced charge density, on a thin conductor, consisting of
the union of $g+1$ disjoint intervals, 
$J:=\cup_{j=1}^{g+1}(a_j,b_j),$
with respect to an external potential. In the context of
random matrix theory this object gives the eigenvalue 
fluctuations of Hermitian random matrix ensembles
where the eigenvalue density is supported on $J.$
\end{abstract}

\bigskip
\noindent
{\bf Running title}: Coulomb Fluid
\vskip .3cm $\dag\;$e-mail: y.chen@ic.ac.uk
\vskip .3cm $\dag\dag\;$e-mail:t.grava@ic.ac.uk 

\bigskip

\noindent
\section{Introduction}
Consider the minimization problem
\begin{equation}
\label{EM}
E=\inf_{\mu\in \mathcal{A}}\left[-\int\int\log|x-t|\mu(x)\mu(t)dxdt+
\int\v(x)\mu(x)dx\right]
\end{equation}
where the set $\mathcal{A}$ consists of all positive 
Lebesgue measures 
$
\mu(t)dt$ such that $\int \mu(t)dt=1$. 
The above formula describes the electrostatic equilibrium in which
charges are placed
on the real line in the presence of an external field $\v(x)$.
For analytic external fields 
it is well known that the infimum (\ref{EM}) is attained at a unique measure
$\s(x)dx$ which is called the induced charge density. 
Moreover, the support 
 of the induced charge density  is generically characterised  
by a finite number of disjoint intervals
$J:=\cup_{j=1}^{g+1}(a_j,b_j)$ \cite{ST}.

At electrostatic equilibrium, the induced 
charged density $\s(x),$  $x\in J$,
satisfies the  following integral equation 
\bea
\label{CP}
\v(x)-2\int_{J}\ln|x-t|\s(t)dt=\A={\rm constant},\;x\in J.
\eea
where the constant $\A$ is the Lagrange multiplier which fixes
the constraint
\bea\label{con}
\int_{J}\s(x)dx=1.
\eea
To determine $\s$, we convert (2) into a singular integral equation
by taking the derivative of (\ref{CP})  w.r.t. $x$, that is
\begin{equation}
\label{sing}
2P\int_{J}\frac{\s(t)dt}{x-t}=\frac{d\v(x)}{dx},\;\;x\in J,
\nonumber
\end{equation}
where $P$ denotes the principal value of the singular  integral.
 For generic potential $\v$ such that $\v^{\prime}(x)$ is H\"older
continuous\footnote{ A
function  $f(x)$  is  H\"older continuous 
if
\[
|f(x_1)-f(x_2)|<c|x_1-x_2|^{\delta},\]
for all 
$x_1,x_2$ in the domain of $f(x)$, for a  constant $c>0$ and  
$0<\delta\leq 1$. }
 the solution, $\s$, of the singular integral equation (\ref{sing}),
which is bounded at the end points of $J$, is necessarily zero 
there \cite{Mus} ,
i.e. 
\[
\s(a_j)=0=\s(b_j),~~~j=1,\dots,g+1.
\]
\noindent
The end points $\{a_j,b_j\}_{j=1}^{g+1}$ of the support of $\sigma$ are determined by
(\ref{con}),
\bea\label{norm}
\int_{b_j}^{a_{j+1}}\s(x)dx=0,\;\;j=1,2,...,g,
\eea 
and by the moment conditions,
\bea
\int_{J}\frac{x^{k}\v^{\pr}(x)}
{{\sqrt {\prod_{j=1}^{g+1}(x-a_j)(x-b_j)}}}dx=0,\;\;\;
k=0,...,g.
\eea
See \cite{Ju}.
The equation (\ref{CP}) also  arises  from a mean-field approach to random
Hermitian matrix ensembles \cite{Dyson1}; $\s(x),$ is the averaged
eigenvalue density, $<\vr(x)>,$ where 
$\vr(x):=\sum_{\nu=1}^{N}\de(x-x_{\nu})/N,$ is the microscopic
density of the eigenvalues, $\{x_{\nu}\}_{\nu=1}^{N},$ of a
$N\times N$ Hermitian random matrix. The validity of the 
mean-field approximation,  for large $N$,
is discussed in \cite{Dyson2}. An easy
calculation shows that a functional derivative \cite{bee} 
of $\s$ w.r.t.
$\v$ is the density-density correlation function{\footnote
{Taking a functional derivative of 
$$
<\vr(x)>:=\frac{\int{\rm e}^
{-(H[\vr]+\int\v(x^{\pr})\vr(x^{\pr})dx^{\pr})}\vr(x)D\vr}
{\int{\rm e}^{-(H[\vr]+\int\v(x^{\pr})\vr(x^{\pr})dx^{\pr})}
D\vr},
$$
w.r.t. $\v(t)$ gives (\ref{corr})}}
:
\bea
\label{corr}
\C_{\vr\vr}(x,t):=\frac{\de \s(x)}{\de\v(t)}=\frac{\de<\vr(x)>}{\de\v(t)}
=<\vr(x)><\vr(t)>-<\vr(x)\vr(t)>,\quad x,t\in J
\eea
and must satisfy the obvious sum rules,
$\int_{J}\C_{\vr\vr}(x,t)dt=0=\int_{J}C_{\vr\vr}(x,t)dx.$
Furthermore, from (\ref{corr})  $\C_{\vr\vr}(x,t)=\C_{\vr\vr}(t,x).$       
In this work we explicitly determine $\C_{\vr\vr}(x,t)$ as a function 
of the end-points of the interval  $J:=\cup_{j=1}^{g+1}(a_j,b_j)$.
It turns out that the density-density correlation function can be
identified with the Bergman kernel of a Riemann surface which is a two-sheeted covering of the complex plane.
\section{Determination of the density-density correlation function}
We fix some notation. 
We consider the hyperelliptic  Riemann surface $\mathcal{S}_g$ of
genus $g$, defined by the equation
$$
\mathcal{S}_g:=\Bigg\{(y,z), \;z\in \Cp^1,~~ y^2=\prod_{j=1}^{g+1}(z-a_j)(x-b_j)\Bigg\}.
$$ 
The projection  $(y,z)\rightarrow z$ defines $\mathcal{S}_g$ as a two-sheeted
covering of the complex plane $\Co$ cut along $J.$ On $\mathcal{S}_g$ we define  the canonical cycles $\{\al_k,\bt_k\}_{k=1}^{g}$  
shown in Figure 1.     
Let
\bea
\U_g(x):=\frac{i}{\pi}\left(x^g+\sum_{j=0}^{g-1}\kappa_jx^j\right),
\eea
with $\kappa_j$, $j=1,...,g,$ determined by 
\bea
\int_{b_j}^{a_{j+1}}\frac{\U_g(x)}{y(x)}dx=0,\;\;j=1,...,g.
\eea
For completeness we first determine $\A$ as a function of $\v$.
Multiply (\ref{CP}) by $\U_g(x)/y(x)$, integrate w.r.t. $x$ over
$J$ and noting that, 
\[
\int_{J}\frac{\U_g(x)}{y(x)}dx=1,
\]
we find
\bea
\mbox{} \A[\v]&=&-2\int_{J}\s(t)dt\int_J
\frac{\U_g(x)}{y(x)}\ln(b_{g+1}-x)dx+
\int_{J}\frac{\v(x)\U_g(x)}{y(x)}dx\label{Av}\nonumber\\
\mbox{} &=&2\V[J]+\int_{J}\frac{\v(x)\U_g(x)}{y(x)}dx,
\eea
where
\bea
\V[J]:=\int_{b_{g+1}}^{\infty}
\left(\frac{\pi}{i}\frac{\U_g(t)}{y(t)}-\frac{1}{t}\right)dt-
\ln b_{g+1}.\eea
The above  integral is along any path connecting $b_{g+1}$ and
$\infty^+$ laying on the upper half plane of the upper sheet. 
So in the absence of the external field, $\A[0]/2$ is entirely
determined by the end points of the conductor. 

Note that 
\bea
\label{a0}
0=\delta\left(\int_{J}\s(x)dx\right)
=\sum_{j=1}^{g+1}
\left(\delta b_j\s(b_j)-\delta a_j\s(a_j)\right)
+\int_{J}\delta\s(x)dx=\int_{J}\de\s(x)dx.
\eea
Using similar calculations,
\begin{equation}
\label{norm0}
\int_{b_j}^{a_{j+1}}\de\s(x)dx=0,~~~~j=1,\dots,g.
\end{equation}
So performing a variation on (\ref{CP}) gives,
\bea
\label{a1}
\de\v(x)-2\sum_{j=1}^{g+1}
\left(\ln|x-b_j|\s(b_j)-\ln|x-a_j|\s(a_j)\right)-
2\int_{J}\ln|x-t|\de\s(t)dt=\de\A.
\eea
Also in this case, we multiply  the above relation by $\U_g(x)/y(x)$,
 integrate w.r.t. $x$ over $J$ and, by (\ref{Av}), we obtain  
\bea
\delta\A=\int_{J}\frac{\U_g(x)}{y(x)}\delta\v(x)dx.
\eea
Taking a derivative w.r.t. $x$ on (\ref{a1}) produces the singular
integral equation
\bea
\label{a2}
2P\int_{J}\frac{\de\s(t)}{x-t}dt=\frac{d\de\v(x)}{dx},\;\;x\in J.
\eea
Now a possible form for $\de\s(x),\;\;x\in J,$ reads,
\bea
\label{a3}
\de\s(x)=\frac{1}{2\pi^2y(x)}P\int_{J}
\frac{y(t)}{t-x}\frac{d\de\v(t)}{dt}dt
-\sum_{k=1}^{g}\frac{\vp_k(x)}{2\pi^2 y(x)}
\int_{J}\frac{d\de\v(t)}{dt}y(t)\left(\int_{\al_k}
\frac{ds}{(t-s)y(s)}\right)dt.
\eea
In order that $\de\s$ actually satisfies 
(\ref{a2}), $\vp_k$ is taken to be a polynomial
of degree $g-1;$
\bea
\vp_k(x)=\sum_{l=1}^{g}\ga_{kl}x^{g-l},\;\;k=1,...,g
\eea
with yet undetermined $\ga_{kl}.$ See \cite{Ga}.
With this choice of $\vp_k$, (\ref{a0}) and (\ref{a2}) are satisfied.
Now $\vp_k(x)dx/y(x)$ is an Abelian differential of the first 
kind on the Riemann surface $\mathcal{S}_g$. If we choose
$\ga_{kl}$ in such way that
\bea
\int_{\al_j}\frac{\vp_k(x)}{y(x)}dx=\de_{kj},
\eea
then (\ref{norm0}) are also satisfied. This completes the solution of
(\ref{a2}). To see that (\ref{norm0}) is satisfied, note that,
\bea
\int_{b_j}^{a_{j+1}}\de\s(x)dx=-\frac{1}{2}
\int_{\delta_j}\de\s(x)dx,\nonumber
\eea
where 
\bea
\de_j&=&\al_j\setminus\al_{j+1},\;\;j=1,...,g-1\nonumber\\
\de_g&=&\al_g.  
\eea
Here the cycles $\de_j,\;\;j=1,..,g$ are shown in Figure 1. 
\begin{figure}[tb]
\centering
\vspace{1in}
\includegraphics[height=1.3in]{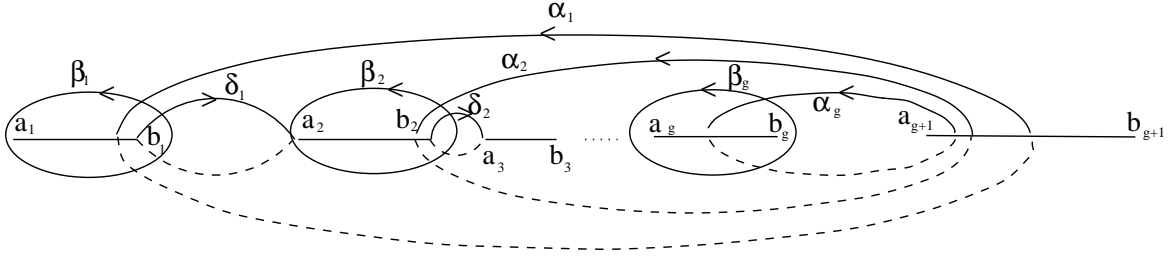}
\caption{The cycles 
$\{\al_{j},\bt_{j},\de_{j}\}_{j=1}^{g}$. The part of the 
cycles that lie in the lower sheet are indicated by broken 
lines.}
\end{figure}
Integrating by parts in (\ref{a3}) and noting that
$y(a_j)=0=y(b_j)$ gives,
\bea
\de\s(x)=P\int_{J}\C_{\vr\vr}(x,t)\de\v(t)dt,\;\;x\in J,
\eea
where
\bea
\label{ddcorr}
\C_{\vr\vr}(x,t):=\frac{\partial}{\partial t}
\left(\frac{y(t)}{2\pi^2y(x)(x-t)}-
\sum_{k=1}^{g}\frac{\vp_k(x)y(t)}{2\pi^2y(x)}
\int_{\al_k}\frac{ds}{(s-t)y(s)}\right),\;\;x,t\in J.
\eea
An inspection of (\ref{ddcorr}) shows that sum rules are satisfied. 
Furthermore, $\C_{\vr\vr}(x,t)$ can be identified with the
Bergman kernel of the Riemann surface $\mathcal{S}_g$ and is therefore 
symmetric under the exchange of $x$ and $t$, namely
$\C_{\vr\vr}(x,t)=\C_{\vr\vr}(t,x)$ \cite[p218]{Schiffer}.

A more direct way to check the symmetry property of $\C_{\vr\vr}(x,t)$
 is shown below.
Let $\pi_j(t)dt/y(t)$ be an Abelian differential of 
the second kind with vanishing $\al$ periods
\bea
\label{normk}
\int_{\al_k}\frac{\pi_j(t)}{y(t)}dt=0,\;\;j,k=1,...,g,
\eea 
and with behavior  at infinity
\bea
\label{asym}
\frac{\pi_j(t)}{y(t)}dt\sim\pm\left(t^{j-1}+{\rm O}(t^{-2})\right)
dt,\;\;\;t\sim \infty^{\pm}.
\eea
The quantity  $\pi_j$  is a polynomial in $t$  of degree $ g+j$: 
\bea
\pi_j(t)=\G_0t^{g+j}+\G_1t^{g+j-1}+...+\G_jt^g
+\a_1t^{g-1}+\a_2t^{g-2}+...+\a_g,
\eea
where the constants $\a_j$'s are determined by (\ref{normk}) and the
   constants $\G_k$'s determined by (\ref{asym}) are the
coefficients of the expansion
\bea
y(z)\sim z^{g+1}\left(\G_0+\frac{\G_1}{z}+\frac{\G_2}{z^2}+...
\right),\;\;z\sim \infty^+.
\eea
From the Riemann bi-linear relations the second term in 
(\ref{ddcorr}) 
can be expressed in terms of $\pi_j(t),$ without involving
the constants $\ga_{jk}$ \cite{Ba},\cite{Grava}. So,
\bea
2\pi^2\C_{\vr\vr}(x,t)=
\frac{y^{\pr}(t)}{y(x)(x-t)}+\frac{y(t)}{y(x)(x-t)^2}   
+\frac{1}{y(x)y(t)}
\sum_{k=1}^{g}x^{g-k}\sum_{j=1}^{k}(2j)\G_{k-j}\pi_j(t).
\eea
The reader can now check that $\C_{\vr\vr}(x,t)-\C_{\vr\vr}(t,x)$
vanishes identically in $x$ and $t.$ For $g=1$ this symmetry
can be established in a straightforward calculation. The kernel,
$\C_{\vr\vr}$ will be used, in a later publication, for
computing the distribution functions of linear statistics
which is of interest in random matrix theory. For a discussion
concerning linear statistics see \cite{Chen}.


\end{document}